# Optimum orientation versus orientation averaging description of cluster radioactivity

W. M. Seif [1], M. Ismail [1], A. I. Refaie [1], and L. H. Amer [1,2]
[1] Cairo University, Faculty of Science, Department of Physics, Giza 12613, Egypt
[2] Taiz University, Faculty of Science, Department of Physics, Taiz, Yemen

**Background**: The deformation of the nuclei involved in the cluster decay of heavy nuclei affect seriously their half-lives against the decay.

**Purpose**: We investigate the description of the different decay stages in both the optimum orientation and the orientation-averaged pictures of the cluster decay process.

**Method**: We consider the decays of $^{232,233,234}$U and $^{236,238}$Pu isotopes. The quantum mechanical knocking frequency and penetration probability based on the Wentzel-Kramers-Brillouin approximation are used to find the decay width.

**Results**: We found that the orientation-averaged decay width is one or two orders of magnitude less than its value along the non-compact optimum orientation. The difference between the two values increases with decreasing the mass number of the emitted cluster. Correspondingly, the extracted preformation probability based on the averaged decay width increases with the same orders of magnitude compared to its value obtained considering the optimum orientation. The cluster preformation probabilities ($S_c$) obtained in the two considered schemes give more or less comparable agreement with the Blendowske–Walliser (BW) formula based on the preformation probability of α ($S_\alpha^{ave.}$) obtained from the orientation-averaging scheme. All the obtained results, including those obtained in the optimum-orientation scheme, deviate substantially from the BW law based on $S_\alpha^{opt.}$ obtained using the optimum-orientation scheme.

**Conclusion**: In order to account for deformations of the participating nuclei, it is more relevant to calculate the decay width by averaging over the different possible orientations of the participating deformed nucleus, rather than considering the corresponding non-compact optimum orientation.

## I. INTRODUCTION

The spontaneous cluster radioactivity is an intermediate phenomenon between alpha decay and spontaneous fission. This rare phenomenon was first discussed theoretically in 1980 by Săndulescu, Poenaru and W. Greiner [1]. The experimental confirmation came four years later with the observation of the $^{14}C$ spontaneous emission from $^{223}Ra$, by Rose and Jones [2]. To date, about 35 cluster decay modes have been observed for nuclei in the trans-lead region of 221<A<242 and 87<Z<96, with daughter nuclei in the neighborhood of $^{208}Pb$. The observed emitted clusters include $^{14}C$ [2], $^{15}N$ [3], $^{18,20}O$ [4,5], $^{22,24,25,26}Ne$ [6–10], $^{23}F$ [11], $^{28,30}Mg$ [6,12,13,14], and $^{32,34}Si$ [13,15]. In principle, These decays are characterized with long partial half-lives lying between $10^{11}$ s and $10^{28}$ s, and very small branching ratios relative to $\alpha$-decay in the range from $10^{-9}$ and $10^{-16}$ [16]. Two more islands of active cluster radioactivity are indicated theoretically [17]. One of them above 100Sn while the second above N=82 [18]. Cluster decays with atomic number A>28 from superheavy elements of Z>110 is theoretically predicted as well [19].

One of the confirmed factors that affect the spontaneous decay process is the deformation of the involved nuclei, in orientation degrees of freedom. The influence of nuclear deformation starts at the early step of preforming the individual clusters inside the radioactive nucleus. After preformation, it affects the interaction potential between the two clusters [20,21]. This appears clearly in the orientation distributions of the depth and width of the pocket in the internal region of the interaction potential, the saddle point, and the height, radius and width of the associated Coulomb barrier [22,23,24]. This impacts the assault frequency, the barrier penetration probability, and consequently the decay width [25,26,27]. A net effect of nuclear deformation on the decay process is observed to decrease the half-life time by several orders of magnitude. Considering optimum orientations for non-compact [28] configuration of decay products, the inclusion of quadrupole deformations if found to decrease the calculated half-lives by 2–7 orders of magnitude [25,29]. Adding the higher-multipole deformations increased the calculated half-lives with the same orders of magnitude in some studies [30], but it affected the results dramatically in other studies [29]. The values of the extracted preformation had been increased by approximately the same orders of magnitudes. This indicates that the uncertainty concerning the calculations involving deformed nuclei still large. Also, the calculated partial widths of proton decays have been employed to predict information on the deformation parameters of the involved nuclei, by comparing them to the observed experimental values [31]. Within the Coulomb and proximity potential model, the computed half-lives of most of the cluster decays of $^{248-254}Cf$ isotopes is roughly 2-6 orders of magnitude smaller [32], for the calculations including deformations. In addition to the static ones, the dynamical surface deformations [33] of both clusters reduce the decay half-life time as well. The dynamical deformation of a cluster is correlated with its ground-state static deformation. The deformation of the daughter nucleus is found to influence the decay half-life time much more than that of the emitted light cluster [33].

In the preformed cluster model of spontaneous cluster decay, both the emitted light particle and the daughter are clustered as individual entities inside the parent nucleus [34], with a certain preformation probability, as a first stage of the decay process. As soon as they form, the light cluster tries to tunnel through the Coulomb barrier between the two formed clusters. It does so in the confining attractive pocket region of the interaction potential. The number of trials on the barrier per unit time is the so called assault frequency. If any of the formed clusters is deformed, or both, we have then orientation distribution of Coulomb barriers, instead of a single one. In this case, the decay takes place in two scenarios. In the first one, the light cluster oscillates and then emits at a certain optimum orientation, with respect to the daughter nucleus, at which the Coulomb barrier distribution has minimum height [29]. This assumption is based on the slow motion of the spontaneous cluster emission. The magnitudes and signs of the multipole deformation components of the involved nuclei determine their optimum orientations for non-compact configuration [28,35]. The second scenario interprets the cluster decay as an orientation-average process [36,37]. In such consideration, the light cluster tries to tunnel through all possible orientations. Thus, we have an orientation distribution of decay widths. The picture of the decay through non-compact optimum orientation is used to describe the decay process in the most studies of cluster decays [25,29,30,32,33,38,39]. Averaging the decay width over different orientation is frequently used in the alpha decay studies [36,37,40].

However, in the present work we focus on the investigation of the cluster decay process in both the optimum orientation and the orientation-averaged descriptions. In the next section, we outline the theoretical approach of investigating the cluster decay process and the cluster preformation probability inside the parent nucleus. The results are presented and discussed in Sec. III. Finally, Sec. IV gives a brief summary and conclusion.

## II. THEORETICAL FORMALISM

Several theoretical models have been proposed to describe α and cluster radioactivity of heavy nuclei, as a quantum-tunneling phenomena. For instance, the semi-microscopic algebraic cluster model [41,42], the generalized density dependent cluster model [43,44], the combined shell and cluster models [31,34], the multistep shell model [45,46], and the preformed cluster model [29,47] have been developed for this purpose in different studies. Also, the numerical and analytical super asymmetric fission model [48,49], multiparticle R-matrix approaches [34,50], the dinuclear system model of cluster radioactivity [51,52], the generalized liquid drop model [53,54], and the universal curves [55,56] have been employed for the same aim.

In most of the above-mentioned models, the half-life ($T_{1/2}$) of a radioactive parent against a specific cluster decay mode is related to the corresponding decay width $\Gamma_c = \hbar \nu_c P_c$ via the relation,

$$T_{1/2} = \frac{\hbar \ln 2}{S_c \Gamma_c}. \qquad (1)$$

$\nu_c$, $P_c$ and $S_c$ represents the tunneling knocking frequency, the penetration probability, and the preformation probability of the emitted particle, respectively. If any of the formed clusters inside the parent nucleus is deformed, we can use the Wentzel-Kramers-Brillouin (WKB) approximation to express the orientation-dependent knocking frequency and penetration probability, respectively, in the form [26,37]

$$\nu(\theta) = T^{-1}(\theta) = \left[ \int_{R_1(\theta)}^{R_2(\theta)} \frac{2\mu}{\hbar k(r,\theta)} dr \right]^{-1}, \qquad (2)$$

and

$$P(\theta) = \exp\left(-2\int_{R_2(\theta)}^{R_3(\theta)} k(r,\theta)dr\right). \quad (3)$$

Here, $\theta$ is the relative orientation angle between the separation vector joining the centers of mass of the two formed clusters ($\vec{r}$) and the symmetry-axis of the participating deformed cluster. $k(r,\theta) = \sqrt{2\mu|V_T(r,\theta) - Q_c|/\hbar^2}$ is the wave number. $\mu = \frac{m_c m_D}{m_c + m_D}$ represents the reduced mass of the emitted cluster ($m_c$)–daughter ($m_D$) system. $Q_c$(MeV) is the released energy in the considered cluster decay. It is calculated from the mass excess [57] of the involved nuclei. $R_{i=1,2,3}$(fm) are the three turning points for the WKB action integrals in Eqs. (2) and (3). They determine the boundaries of the internal pocket and the Coulomb barrier regions in the interaction potential at which $V_T(r,\theta)|_{r=R_i(\theta)} = Q_c$. The total interaction potential ($V_T(r,\theta)$) between the emitted cluster and the daughter nucleus is usually taken as the sum of the nuclear ($V_N(r,\theta)$), Coulomb ($V_C(r,\theta)$) and the centrifugal ($V_\ell(r)$) potential parts,

$$V_T(r,\theta) = \lambda V_N(r,\theta) + V_C(r,\theta) + V_\ell(r). \quad (4)$$

Here, the normalization factor $\lambda$ of the nuclear part of the interaction potential is determined by applying the Bohr-Sommerfeld quantization condition [58],

$$\int_{R_1(\theta)}^{R_2(\theta)} k(r,\theta)dr = (2n+1)\frac{\pi}{2}. \quad (5)$$

The quantum number $n$ represents the number of nodes of the quasibound radial wave function of the cluster-daughter system [59]. $n$ depends on the ground states of the involved nuclei. For even(Z)-even(N) nuclei, the decaying state would be a zero-nodes eigenstate as interpreted microscopically for $\alpha$-decay in Ref. [60].

The first two turning points ($R_{1,2}(\theta)$) depend on the orientation of the deformed cluster [61]. In the tail region of the interaction potential between an emitted cluster of charge numbers $Z_c$ and daughter nucleus ($Z_D$), the orientation-independent $R_3$ reads,

$$R_3 = \frac{0.72\, Z_c Z_D}{Q_c} + \sqrt{\left(\frac{0.72\, Z_c Z_D}{Q_c}\right)^2 + \frac{\ell(\ell+1)\hbar^2}{2\mu Q_c}}.$$

In this equation, $\ell$ represents the angular momentum carried out by the emitted light cluster to conserve the spin and parity for the considered decay mode. If the emitted cluster is an even-even nucleus, its ground state spin-parity is then $0^+$. In this case, according to conservation laws of spin and parity, the transferred angular momentum by the emitted cluster must satisfy the conditions, $|J_P - J_D| \leq \ell \leq |J_P + J_D|$ and $\pi_P = \pi_D(-1)^\ell$. Here, $J^\pi_{P(D)}$ are the spin(J)-Parity($\pi$) assigned for the involved states of parent (P) and daughter (D) nuclei. Following the principle of least action, the emitted light cluster is assumed to carry out the minimum value of $\ell$ satisfying the conservation rules of spin and parity. In terms of $\ell$, the centrifugal potential part is given as, $V_\ell(r) = \ell(\ell+1)\hbar^2/2\mu r^2$.

Consider the Hamiltonian energy density approach [62], in the frozen density approximation. We can calculate the real nuclear part of the interaction potential between an emitted spherical cluster and a deformed daughter nucleus, $V_N(r,\theta)$ in Eq. (4), via [63-65]

$$V_N(r,\theta) = \int \{H[\rho_{pc}(\vec{x}) + \rho_{pD}(r,\vec{x},\theta), \rho_{nc}(\vec{x}) + \rho_{nD}(r,\vec{x},\theta)] - H_c[\rho_{pc}(\vec{x}), \rho_{nc}(\vec{x})] - H_D[\rho_{pD}(\vec{x}), \rho_{nD}(\vec{x})]\} \, d\vec{x} \,. \tag{6}$$

$H$, $H_c$ and $H_D$ are the Skyrme energy density functional for the whole system, emitted cluster, and daughter nucleus, respectively. $\rho_{ic(iD)}(i = p, n)$ represent the protons ($p$) and neutrons ($n$) density distributions of the emitted cluster ($c$) and the daughter nucleus ($D$). These density functionals are given in terms of the matter densities ($\rho_i(i = p, n)$), the kinetic energy densities ($\tau_i$) and the spin-orbit densities ($\vec{J}_i$) of protons and neutrons, $H = (\hbar^2/2m) \sum_{i=n,p} \left(\tau_i + H_{Sky}\left(\rho_i, \tau_i, \vec{J}_i\right)\right)$. More details about the method of calculations are outlined in Refs. [37,37,65]. In the present calculations we used the Skyrme-SLy4 parameterization [66] for the effective nucleon-nucleon interaction.

Regarding the Coulomb potential ($V_C(r,\theta)$), the direct and exchange parts of the Coulomb density functionals read

$$H_{Coul}(\rho_p) = H_C^{dir}(\rho_p) + H_C^{exch}(\rho_p) = \frac{e^2}{2} \rho_p(\vec{r}) \int \frac{\rho_p(\vec{r}\,')}{|\vec{r} - \vec{r}\,'|} d\vec{r}\,' - \frac{3e^2}{4} \left(\frac{3}{\pi}\right)^{1/3} \left(\rho_p(\vec{r})\right)^{4/3}. \tag{7}$$

The Slater approximation [67] has been used to express $H_C^{exch}(\rho_p)$. To deal with the finite range of the Coulomb force in presence of deformed nuclei, we employed the multipole expansion method to compute the direct part of the Coulomb potential [24,68].

The density distributions of $\alpha$ and spherical nuclei are obtained using self-consistent Hartree-Fock calculations [69] based on the Skyrme-SLy4 interaction. The neutron (proton) density distributions of the participating deformed nuclei are represented by the two-parameter Fermi distribution form,

$$\rho_{n(p)}(r,\theta) = \rho_{0n(p)}\left(1 + e^{(r - R_{n(p)}(\theta))/a_{n(p)}}\right)^{-1}, \qquad (8(a))$$

with the half-density radii (in *fm*)

$$R_{n(p)}(\theta) = R_{0n(p)} \left[1 + \sum_{i=2,3,4,6} \beta_i Y_{i0}(\theta)\right], \qquad (8(b))$$

where

$$\begin{aligned} R_{0n}(fm) &= 0.953\, N^{1/3} + 0.015\, Z + 0.774 \\ R_{0p}(fm) &= 1.322\, Z^{1/3} + 0.007\, N + 0.022 \\ a_n(fm) &= 0.446 + 0.072\, (N/Z) \\ a_p(fm) &= 0.449 + 0.071\, (Z/N). \end{aligned} \qquad (8(c))$$

The radius, $R_{0n(p)}$, and the diffuseness, $a$, parameters given by Eq. (8(c)) are obtained [70] from a fit to the Hartree-Fock calculations of the density distributions, based on the Skyrme-SLy4 interaction. While N and Z are, respectively, the neutron and proton numbers for the considered nucleus, $\beta_i (i = 2,3,4,6)$ represent its multipole deformation parameters [71]. $\rho_{0n(p)}$ are given by the normalization, $\int \rho_{n(p)}(r,\theta)\ d\vec{r} = N(Z)$.

For the sake of investigating, we consider two schemes to find the decay width, and consequently the half-life. In the first one, the decay width is calculated at the optimum orientation of the deformed nucleus for the decay process,

$$\Gamma = \hbar\nu(\theta_{opt})P(\theta_{opt}). \qquad (9)$$

In the second scheme, averaging over all orientations is carried out to find the average decay width

$$\Gamma = \frac{\hbar}{2}\int_0^{\pi} \nu(\theta)\,P(\theta)\,sin\,\theta\ \ d\theta. \qquad (10)$$

We can debrief the preformation probability of the emitted cluster in the parent nucleus from the experimental half-life and the calculated decay width by,

$$S = \frac{ln\,2}{\Gamma\,T_{1/2}^{exp}}. \qquad (11)$$

A phenomenological relation is suggested by Blendowske and Walliser [72,73] to relate the preformation probability of a cluster ($S_c$) of mass number ($A_c$) inside a given heavy nucleus to the preformation probability of the α-cluster ($S_\alpha$) inside it as,

$$S_c = (S_\alpha)^{\frac{A_c-1}{3}}. \qquad (12)$$

This relation is assumed to be valid up to number of $A_c$=28. Taking into account the shell closures in the parent nucleus ($Z_0$, $N_0$), the nucleon paring ($a_p$), and the transferred angular momentum ($a_\ell$) influences, an empirical formula is proposed [37,61] to give $S_\alpha$ as,

$$S_\alpha = \frac{Ae^{-\alpha(Z-Z_0-Z_c)^2}e^{-\beta(N-N_0-N_c)^2} - a_p}{a_\ell}.$$

The description of the quantities and parameters appearing in this relation is given in Ref. [37].

## III. RESULTS AND DISCUSSION

We carried out the calculations for α and heavier cluster decays of $^{232,233,234}$U, and $^{236,238}$Pu radioactive isotopes. The considered decays of U isotopes are $^{232}$U($^{228}$Th($\beta_2$=0.158,$\beta_4$=0.076)),α), $^{232}$U($^{208}$Pb,$^{24}$Ne($\beta_2$=0.191,$\beta_4$=0.024)), $^{232}$U($^{204}$Hg,$^{28}$Mg ($\beta_2$=0.254,$\beta_4$= –0.022)), $^{233}$U($^{229}$Th($\beta_2$=0.190,$\beta_4$= 0.114,$\beta_6$= 0.020),α), $^{233}$U($^{209}$Pb,$^{24}$Ne), $^{233}$U($^{205}$Hg, $^{28}$Mg), $^{234}$U ($^{230}$Th($\beta_2$=0.185,$\beta_4$= 0.092),α), $^{234}$U($^{210}$Pb,$^{24}$Ne), $^{234}$U($^{208}$Pb, $^{26}$Ne), and $^{234}$U($^{206}$Hg,$^{28}$Mg). The considered decays of Pu isotopes are $^{236}$Pu($^{232}$U ($\beta_2$=0.201, $\beta_4$= 0.099),α), $^{236}$Pu($^{208}$Pb,$^{28}$Mg), $^{238}$Pu($^{234}$U($\beta_2$=0.220,$\beta_4$= 0.102,α), $^{238}$Pu($^{210}$Pb,$^{28}$Mg),

$^{238}Pu(^{208}Pb, ^{30}Mg(\beta_2=0.170, \beta_4=-0.012)$, $^{238}Pu(^{206}Hg, ^{32}Si(\beta_2=-0.144, \beta_4=-0.010))$. Except for the odd-A nucleus $^{229}Th$ [74], the deformation parameters for the involved nuclei were taken from Ref. [71]. Only the decay $^{234}U$ ($^{208}Pb$, $^{26}Ne$) has no deformed nuclei involved. However, the participating deformed nuclei possess only deformations of reflection symmetry ($\beta_2$, $\beta_4$, $\beta_6$). Thus, we performed the calculations along the orientation angles $\theta = 0^0 - 90^o$, with respect to the symmetry axis of the deformed nucleus. The nuclear part of the interaction potential ($V_N(r,\theta)$) is calculated in the framework of the energy density formalism, Eq. (6), based on the Skyrme-SLy4 nucleon-nucleon interaction. Advantages of this interaction are to include the pairing and shell effect influences in the calculations [61]. After computing the Coulomb potential, and the centrifugal potential for the studied decays of $^{233}U$, the obtained total potential (Eq. (4)) has been implemented to find the penetration probability (Eq. (3)) and knocking frequency (Eq.(2)) along different orientations. We employed the decay width obtained along the optimum orientation of the considered decay mode (Eq, (9)) and the orientation averaged one (Eq. (10)) to deduce the preformation probability (Eq. (11)) by the two schemes.

The optimum orientation for all the considered decays involving prolate nuclei is $\theta_{opt}=0^o$. For the decay of $^{238}Pu$ ($^{206}Hg$,$^{32}Si$) which involves an oblate $^{32}Si$ nucleus, the optimum orientation is $\theta_{opt}=90^o$. Displayed in Fig. 1 are the calculated decay widths for the considered decays. We compare in this Figure between the decay widths calculated along the optimum orientations and those obtained by averaging over different orientations. Generally, the presented results show that the average decay width is about one or two orders of magnitude less than it in the optimum orientation. The maximum difference between the two values is obtained for the α-decays. This difference decreases with increasing the mass number of the emitted cluster. Typical one or two orders of magnitude appear as increasing in the extracted preformation probability based on the averaged decay width with respect to its value based on the optimum orientation calculations. The extracted values of the cluster preformation probability are presented in Table I.

The first three columns in Table I identify, respectively, the parent, daughter and the emitted cluster participating in the considered decays. The experimental released energy, Q(MeV) [57], and the experimental half-lives, $T_{1/2}^{exp}$ (s) [75], used in the calculations are shown in columns 4 and 5, respectively. Presented in the sixth column of Table I are the deduced values of the preformation probability for the considered decay modes, based on the optimum orientation calculations. The same quantity obtained from the orientation-average calculations is presented in the eighth column. Also shown in Table I are the estimated cluster preformation probabilities using the phenomenological formula given by Eq.(12), based on the $\alpha$ preformation probability obtained from the optimum orientation (column 7) and the orientation-average (column 9) calculations. As seen in Table I, considering the optimum orientation yields preformation probability of $\alpha$-cluster inside the presented parent nuclei in the order of $10^{-3}$ and $10^{-4}$. The orientation-average calculations yield $\alpha$ preformation probability in a larger order of $10^{-2}$. For the same considered radioactive isotopes, most of the semi-microscopic spherical and deformed calculations of $\alpha$ decay yield $\alpha$ preformation probability of the order of $10^{-1}$–$10^{-2}$ [36, 76-78]. To evaluate the reliability of these obtained values we plotted Fig. 2.

In Fig. 2, we try to assess the correlation between the obtained preformation probabilities of heavy clusters to the one of $\alpha$-cluster inside the same parent nucleus, based on the two considered schemes. We present in Fig. 2 the estimated preformation probabilities for the $\alpha$ and heavier clusters inside $^{232}$U (Fig. 2(a)), $^{233}$U (Fig. 2(b)), $^{234}$U (Fig. 2(c)), $^{236}$Pu (Fig. 2(d)) and $^{238}$Pu (Fig. 2(e)), versus the mass numbers of the formed light clusters ($A_c$). The preformation probabilities from both the optimum orientation (open symbols) and the orientation-average (solid symbols) schemes are displayed in Fig. 2. The Blendowske–Walliser dependence of the cluster preformation probability (Eq. (12)) based on both the extracted $S_\alpha^{opt.}$ (dashed lines) and $S_\alpha^{ave.}$ (solid lines) is also shown in Fig. 2. For comparison, we also added results from Ref. [29] in the same figure (open triangles). As seen in Fig. 2, the preformation probabilities obtained using the orientation-averaged scheme ($S_c^{ave.}$) for the clusters heavier than $\alpha$ particle are close to the empirical relation of Blendowske–Walliser (BW) based on the preformation probability of $\alpha$ ($S_\alpha^{ave.}$) inside the same isotope. The cluster preformation probabilities based on the optimum orientation scheme ($S_c^{opt.}$) are within one or two orders of magnitude of the BW formula based on $S_\alpha^{ave.}$. Moreover, except for the $^{30}$Mg decay of $^{238}$Pu, the cluster preformation probabilities obtained in Ref. [29] based on the optimum orientation scheme are also within three orders of magnitude of the BW dependence based on $S_\alpha^{ave.}$. However, all the cluster preformation probabilities obtained in the two considered schemes give more or less comparable agreement with the BW formula based on $S_\alpha^{ave.}$. On the other hand, all the obtained results, including those obtained in the optimum orientation scheme, deviate substantially from the BW law based on the preformation probability of $\alpha$ given using the optimum scheme ($S_\alpha^{opt.}$). The deviation from the BW formula based on $S_\alpha^{opt.}$ reaches extremely large values of sixteen (seventeen) orders of magnitude for the optimum orientation (orientation-averaged) results. This deviation increases considerably with increasing the mass number of the emitted cluster. However, this makes the optimum orientation scheme, used to describe the $\alpha$ and cluster radioactivity, very questionable.

## IV. SUMMARY AND CONCLUSION

We investigated the predicted $\alpha$ and heavy-cluster decays of $^{232,233,234}$U and $^{236,238}$Pu radioactive nuclei in the framework of the preformed cluster model. The deformations of the involved nuclei are taken into account. We used the energy density formalism based on the Skyrme-SLy4 nucleon-nucleon interaction to obtain the nuclear part of the interaction potential between the emitted cluster and the daughter nucleus. Along different orientations of the participating deformed nucleus, we calculated the decay width by computing the WKB penetration probability and knocking frequency. We estimated the cluster preformation probability using the experimental half-life and the calculated decay width in both the non-compact optimum-orientation and the orientation-averaged schemes.

We found that the difference between the orientation-averaged decay width and its value along the optimum orientation amounts to about one or two orders of magnitude. The average decay width is less. Consequently, this appears as increasing in the estimated cluster preformation probability based on the orientation-averaged calculations, with the same typical difference. This difference increases with decreasing the mass number of the emitted cluster ($A_c$). The cluster preformation probabilities obtained by both schemes show comparable agreement with the Blendowske–Walliser formula based on $S_\alpha^{ave.}$ obtained by the orientation-averaging scheme. All the obtained preformation values, including those obtained in the optimum orientation scheme, deviate substantially from the BW law based on $S_\alpha^{opt.}$. The deviation increases considerably with increasing $A_c$. We conclude that the decay width should be calculated by averaging over all possible orientations of the participating deformed nuclei, rather than by considering the non-compact optimum-orientation for the decay process.

## <u>Figures and Tables captions</u>:

**<u>Fig. 1</u>**: The calculated decay widths Γ(MeV) for α and heavier-clusters decays of $^{232,233,234}$U and $^{236,238}$Pu radioactive isotopes, based on both the optimum-orientation and the orientation-averaging schemes.

**<u>Fig. 2</u>**: Comparison between the extracted values of the preformation probability of α ( $S_\alpha$ ) and heavier clusters ($S_c$) inside (a) $^{232}$U, (b) $^{233}$U, (c) $^{234}$U, (d) $^{236}$Pu, and (e) $^{238}$Pu from both the optimum-orientation (Opt. Ori.) and the orientation-average (Ori. Ave.) schemes. The preformation probability is plotted versus the mass number of the emitted cluster ($A_c$). The Blendowske–Walliser dependencies of $S_c$ (Eq. (12)) based on both $S_\alpha^{opt.}$ and $S_\alpha^{ave.}$ are also presented. The results based on the optimum orientation scheme from Ref. [29] are shown for comparison.

**Table I**. The deduced cluster preformation probability, $S_c^{exp}$ (Eq. (11)), for the listed radioactive isotopes (column 1), based on the experimental half-lives $T_{1/2}^{exp}$ (s) (column 5) [75] and the calculated decay width. The calculations are performed by considering the optimum orientation of the deformed nucleus (column 6) and by taking the orientation-average of the decay width over all orientations of the deformed nucleus (column 8). The WKB penetration probability and knocking frequency, with an interaction potential based on Skyrme-SLy4 NN interaction, are used to compute the decay width. The second, third and fourth columns identify, respectively, the daughter nucleus, the emitted cluster and the Q-value of the decay process [57]. Columns 7 and 9 exhibit the preformation probability from the formula given by Eq.(12), based on the $\alpha$ preformation probability obtained from the optimum orientation and the orientation-average calculations, respectively.

**Table I**.

| Parent | Daughter | Emitted Cluster | $Q$ (MeV) | $T_{1/2}^{exp}$ (s) | Optimum orientation | | Orientation-average | |
|---|---|---|---|---|---|---|---|---|
| | | | | | $S_c^{exp}$ (Eq. (11)) | $S_c$ $(Eq. (12))$ | $S_c^{exp}$ (Eq. (11)) | $S_c$ $(Eq. (12))$ |
| $^{232}$U | $^{228}$Th | $^{4}$He | 5.414 | 2.174x10$^{9}$ | 1.70x10$^{-3}$ | | 2.96x10$^{-2}$ | |
| $^{232}$U | $^{208}$Pb | $^{24}$Ne | 62.311 | 2.443 x 10$^{20}$ | 6.08x10$^{-14}$ | 5.49x10$^{-22}$ | 5.94x10$^{-13}$ | 1.93x10$^{-12}$ |
| $^{232}$U | $^{204}$Hg | $^{28}$Mg | 74.320 | >4.348x10$^{22}$ | <2.69x10$^{-14}$ | 1.10x10$^{-25}$ | <2.22x10$^{-13}$ | 1.77x10$^{-14}$ |
| $^{233}$U | $^{229}$Th | $^{4}$He | 4.909 | 5.024x10$^{12}$ | 2.83x10$^{-4}$ | | 1.08x10$^{-2}$ | |
| $^{233}$U | $^{209}$Pb | $^{24}$Ne | 60.486 | 6.978x10$^{24}$ | 3.46x10$^{-16}$ | 6.27x10$^{-28}$ | 6.72x10$^{-15}$ | 8.37x10$^{-16}$ |
| $^{233}$U | $^{205}$Hg | $^{28}$Mg | 74.226 | >3.865x10$^{27}$ | <3.78x10$^{-19}$ | 1.16x10$^{-32}$ | <3.18x10$^{-18}$ | 2.00x10$^{-18}$ |
| $^{234}$U | $^{230}$Th | $^{4}$He | 4.858 | 7.747x10$^{12}$ | 7.73x10$^{-4}$ | | 1.96x10$^{-2}$ | |
| $^{234}$U | $^{210}$Pb | $^{24}$Ne | 58.826 | 8.608x10$^{25}$ | 1.01x10$^{-14}$ | 1.39x10$^{-24}$ | 1.24x10$^{-13}$ | 8.14x10$^{-14}$ |
| $^{234}$U | $^{208}$Pb | $^{26}$Ne | 59.416 | 8.608x10$^{25}$ | 8.94x10$^{-13}$ | 1.17x10$^{-26}$ | | 5.92x10$^{-15}$ |
| $^{234}$U | $^{206}$Hg | $^{28}$Mg | 74.111 | 5.534x10$^{25}$ | 2.89x10$^{-17}$ | 9.83x10$^{-29}$ | 2.39x10$^{-16}$ | 4.31x10$^{-16}$ |
| $^{236}$Pu | $^{232}$U | $^{4}$He | 5.867 | 9.019x10$^{7}$ | 1.05x10$^{-3}$ | | 2.83x10$^{-2}$ | |
| $^{236}$Pu | $^{208}$Pb | $^{28}$Mg | 79.670 | 4.509x10$^{21}$ | 1.36x10$^{-16}$ | 1.52x10$^{-27}$ | 1.10x10$^{-15}$ | 1.17x10$^{-14}$ |
| $^{238}$Pu | $^{234}$U | $^{4}$He | 5.593 | 2.768x10$^{9}$ | 6.87x10$^{-4}$ | | 2.17x10$^{-2}$ | |
| $^{238}$Pu | $^{210}$Pb | $^{28}$Mg | 75.912 | 4.613x10$^{25}$ | 2.97x10$^{-16}$ | 3.39x10$^{-29}$ | 2.46x10$^{-15}$ | 1.07x10$^{-15}$ |
| $^{238}$Pu | $^{208}$Pb | $^{30}$Mg | 76.797 | 4.613x10$^{25}$ | 3.65x10$^{-16}$ | 2.64x10$^{-31}$ | 2.47x10$^{-15}$ | 8.32x10$^{-17}$ |
| $^{238}$Pu | $^{206}$Hg | $^{32}$Si | 91.188 | 1.977x10$^{25}$ | 3.56x10$^{-17}$ | 2.05x10$^{-33}$ | 1.89x10$^{-16}$ | 6.48x10$^{-18}$ |

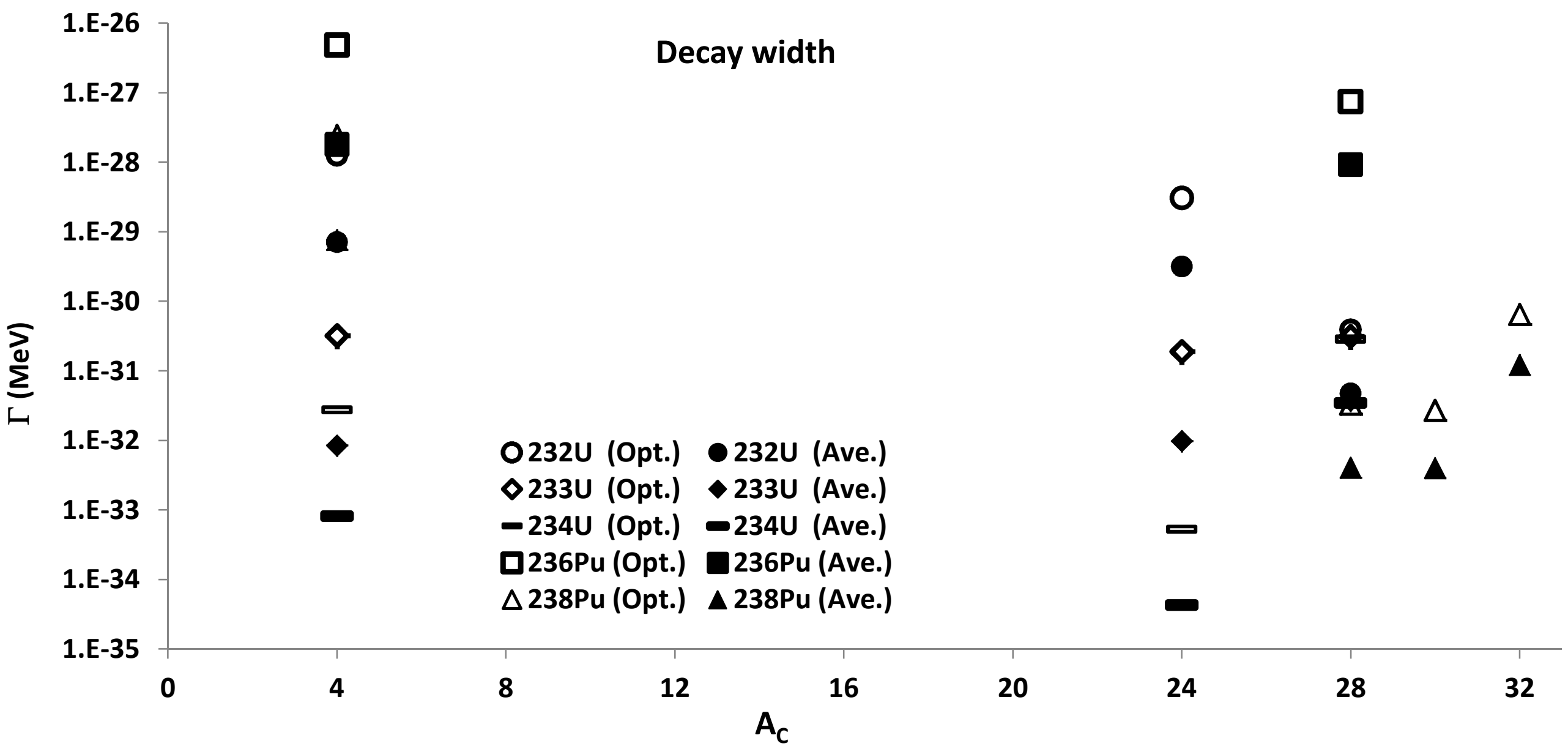

Fig. 1

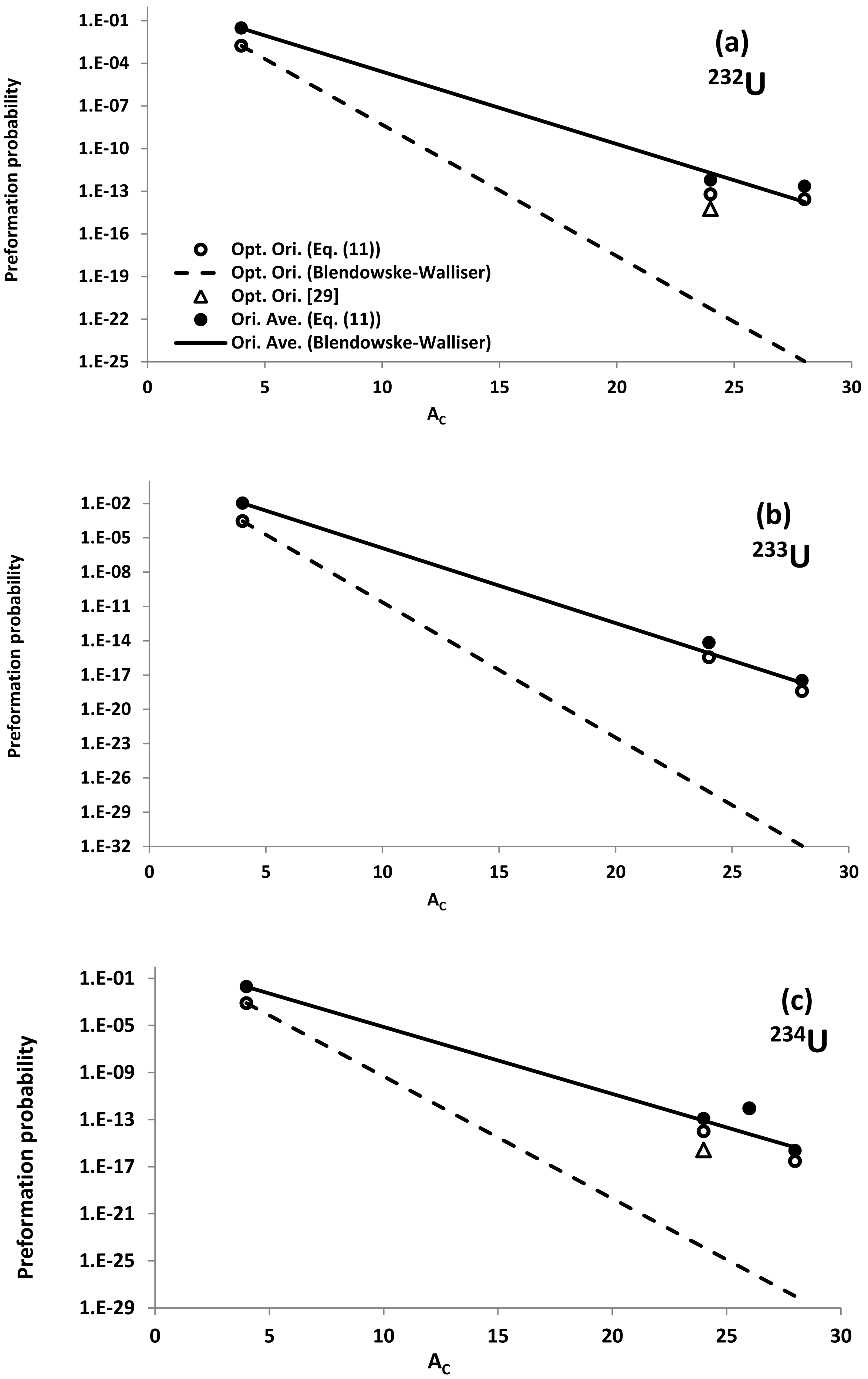

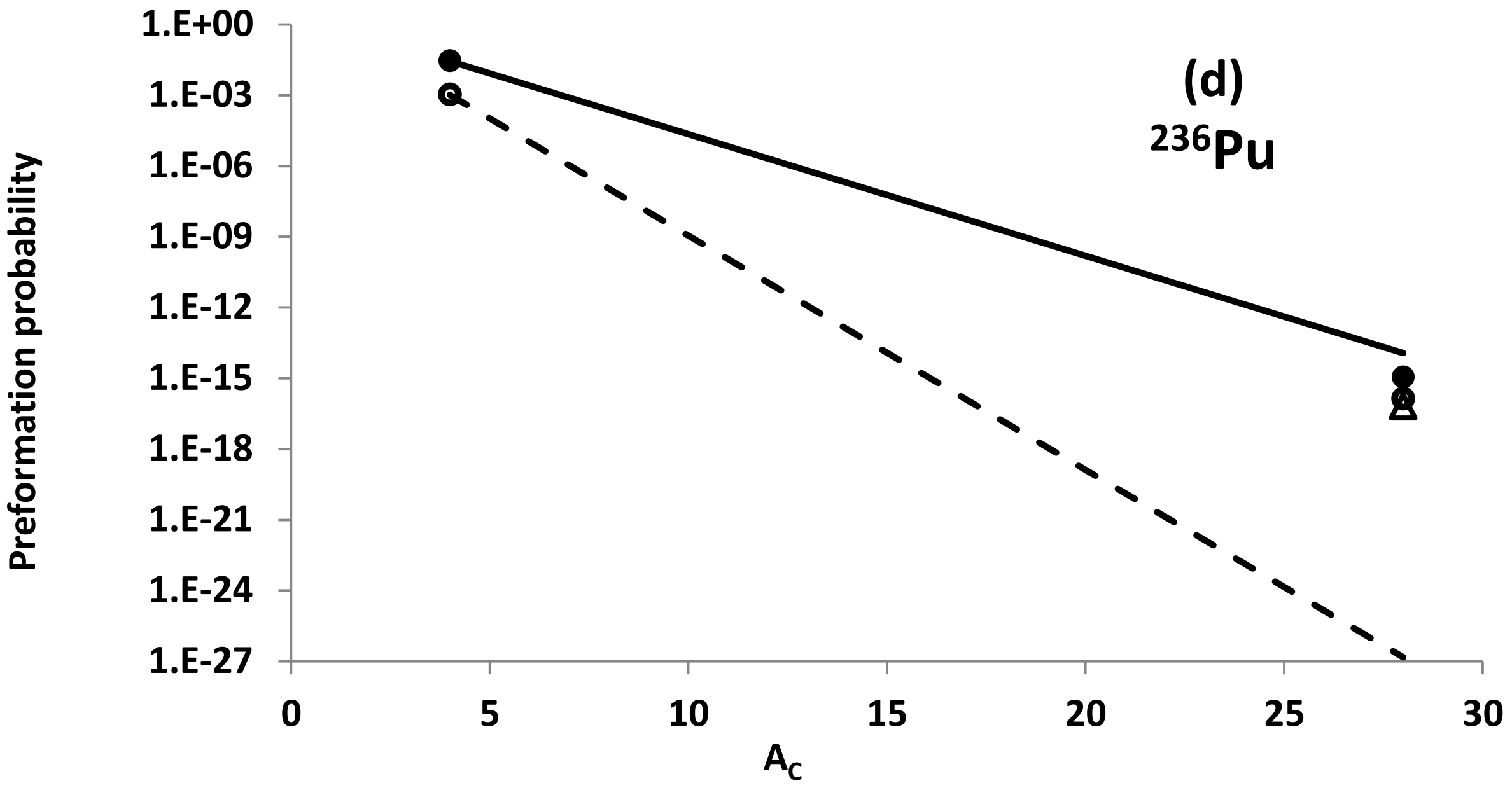

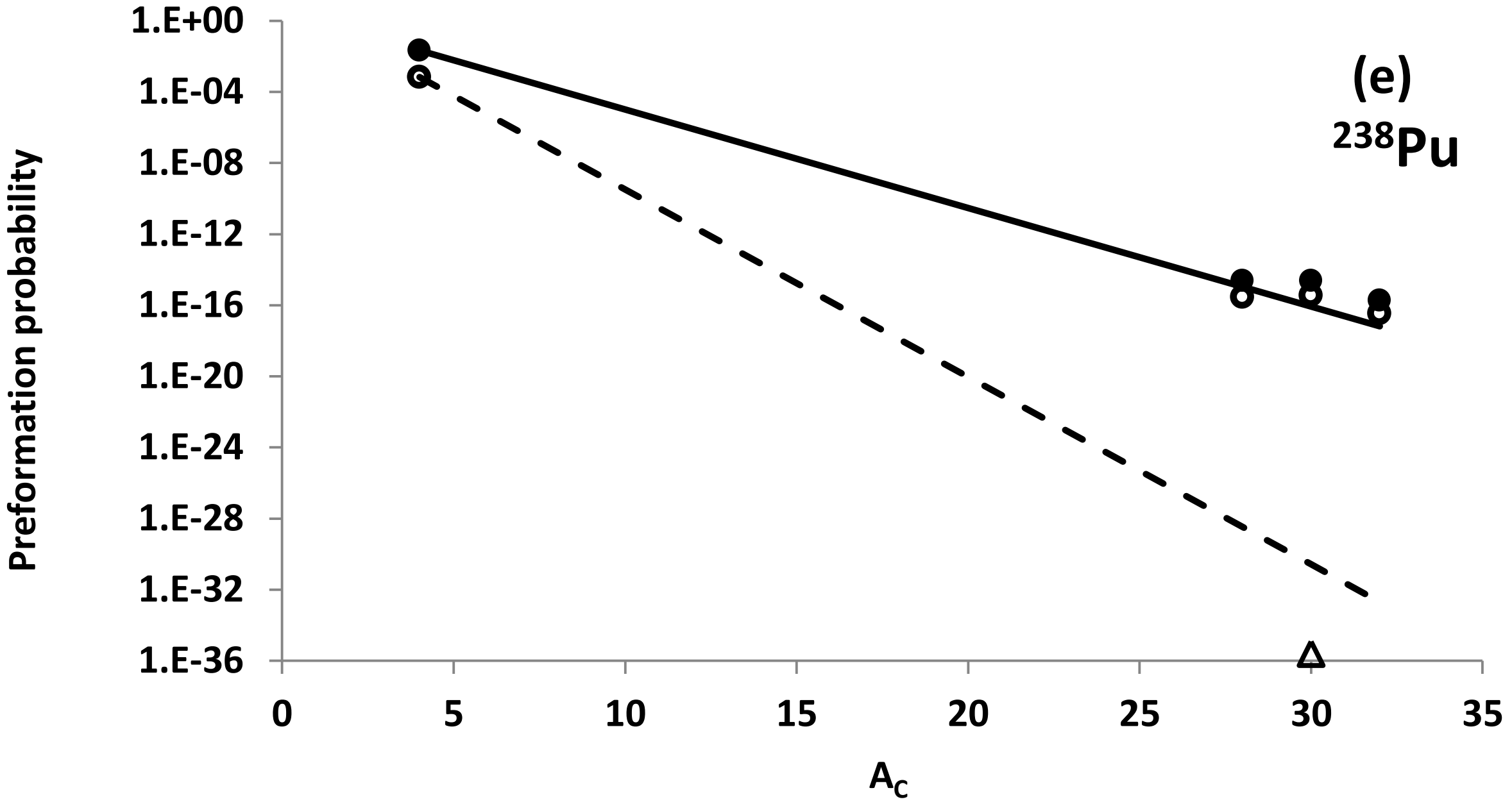